On the Importance of Engaging Students in Crafting Definitions
Angela Little[1] and Leslie Atkins Elliott[2]
[1]Lyman Briggs College, Michigan State University [2]Department of Curriculum, Instruction and Foundational Studies, Boise State University

In this paper we describe an activity for engaging students in crafting definitions. We explore the strengths of this particular activity as well as the broader implications of engaging students in crafting definitions more generally.

## Introduction

Is a hot dog a *sandwich?* The process of trying to articulate a kind of thing, to create a definition, isn't easy. We might have a general sense that some examples clearly fit, but the process of putting this sense into concrete and useful words involves concerted effort.

Often we don't do the explicit work of defining until definitions have real consequences. "Simultaneity," for example, was not an operationally defined term until Einstein's theories hinged on constructing a definition. And so it is key to ask, "for what purpose" is a definition? When New York City decided to create a definition of *sandwich* for the purpose of taxation, they were faced with the difficult task of articulating what would count. Presumably there were reasons to make the category as expansive as possible and hot dogs made the cut[1]:

*Sandwiches include cold and hot sandwiches of every kind that are prepared and ready to be eaten, whether made on bread, on bagels, on rolls, in pitas, in wraps, or otherwise, and regardless of the filling or number of layers. A sandwich can be as simple as a buttered bagel or roll, or as elaborate as a six-foot, toasted submarine sandwich.*

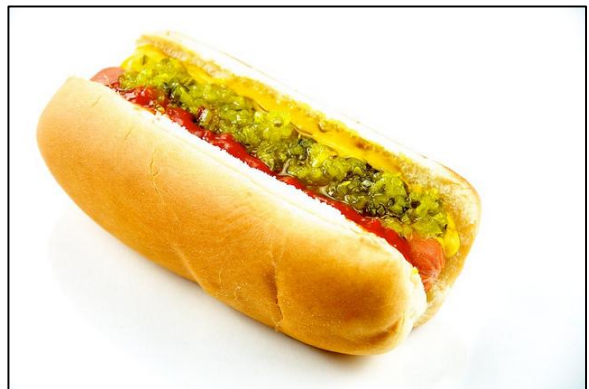

*CC-BY-2.0, Flickr User TheCulinaryGeek, "hot dog."*

**Figure 1.** New York City considers this hotdog to be a sandwich.[2]

## Definitions in Math and Science

Definitional debates are central to math and science research. What if students could be given the experience of crafting a definition in physics? What might they learn from doing so?

Going back as far as the 1930s, some mathematics educators have given their students experiences in constructing and debating definitions. Alan Schoenfeld, in his book *Learning to Think Mathematically*, describes early writing (1938) by a high school teacher, Fawcett, on the importance of having discussions with his students about definitions more generally[3]:

*In baseball, for example, there might be varying definitions of "foul ball" (is a fly ball that hits the foul pole fair or foul?) -- but once one sets the rules, the game can be played with consistency. After such [classroom] discussions [about baseball], Fawcett notes "[n]o difficulty was met in leading the pupils to realize that these rules were nothing more than agreements which a group of interested people had made and that they implied certain conclusions" (p.33). In the mathematical domain, he had his students debate the nature and usefulness of various definitions. Rather than provide the definition of "adjacent angle," for example, he asked the class to propose and defend various definitions.*

Schoenfeld argues that activities like the one described in the quote have the potential to impact students' beliefs about what mathematicians do in practice: "students abstract their beliefs about formal mathematics -- their sense of the discipline -- in large measure from their experiences in the classroom." He goes onto argue for the importance of showing students a sense of mathematics as a "dynamic discipline of sense-making." A similar argument holds for physics.

Yet, in most physics courses students do not spend extended time crafting definitions. Instead, students are given definitions as fully-formed tools to employ in their problem solving or experimental lab work. Some of the day-to-day practice of science *does* include being able to take someone else's definition, make sense of it, and apply it to a new problem. And it is important for students to become skilled in this practice. However, scientists also engage in creating, refining, and debating definitions. Bazerman describes the way that definitions have historically evolved: "In Bacon's day the word acid meant only sour-tasting; then it came to mean a sour-tasting substance; then, a substance which reddens litmus; then, a compound that dissociates in aqueous solution to produce hydrogen ions; then, a compound or ion that can give protons to other substances; and most recently, a molecule or ion that can combine with another by forming a covalent bond with two electrons of the other... The tasting and taster vanish as the structure emerges."[4] As progress was made on understanding acids, the definition shifted from direct human perception ("sour tasting") to greater precision that ultimately relied on theoretical models of the atom.

We describe an activity where students build definitions from scratch so that they can see these earlier aspects of the scientific defining process first hand. While work in constructing definitions can be integrated into a physics class throughout the year[4,5], this activity represents a workshop-like structure that can be used in a range of settings.

**Where to Start**
We propose that a short (1-2 hour) activity where students are asked to define *threshold* is an ideal starting point for many different kinds of physics courses and student backgrounds.[6] We have provided a link to a full lesson plan.[7]

Briefly, the idea is that a physics class is shown a set of examples: bending a pencil until it breaks, blowing up a balloon until it pops, water heated in a tea kettle until it whistles, and pushing a block over the edge of a table until it falls. Students are asked to brainstorm and share out additional examples. Small groups of students then work together to define *threshold* for the purpose of helping younger kids decide what counts.[8]

A benefit of *threshold* is that, unlike a concept such as *force*, there is no easy right answer to look up. Students come at defining from many directions and real debates are possible about examples at the borderlines, e.g. sneezing or milk going bad in the fridge. Like early definitions of *acid* that drew on the human ability to taste, students have the opportunity to consider examples based on their own perception: sudden loud noises, food that goes from good to bad tasting, and visual transformations. They also get to consider the limitations of their perception: some transition timescales are particularly long or short and thus more difficult for human perception to make sense of. Students frequently start with a definition that includes the word "sudden", e.g. "sudden reaction," but often end up considering the role of human perception in judgements of suddenness. For instance, is milk going bad in the fridge "sudden"? Questions around instantaneity arise, sometimes compelling students to bring calculus into their definitions.

The *threshold* activity brings students genuinely into the development a definition for a particular purpose. The activity also allows students to build on resources[9] they have in clarifying and debating language, something they might not always see as relevant to physics.

**Potential Impact on Student Beliefs: Physics Concepts as Connected**
Students in physics courses often develop beliefs that physics is a disconnected set of facts[10,11]. The *threshold* activity is intended to push back on this idea by emphasizing connections. We provide an existence proof example of connections made possible through this activity. One small group of college physics students that author Little worked with explored an idea that thresholds occur at a point where two curves cross one another (see Figure 2). The group decided to check this idea of "crossing" with two specific examples: water boiling and 'a block falling off the edge of a table.' They attempted to graph the physical quantities in both examples to see if they could create a similar looking graph.

How often is it the case that physics students see connections between mechanics and thermodynamics in this way? Yet, in our experience facilitating this definitional activity, such connections were not atypical. Not only do students see connections within physics, they see connections across STEM and beyond. One small group that Little worked with discussed examples as varied as wolf-sheep predation, the stock market, eating too many cookies until you get sick, and human anger and frustration.

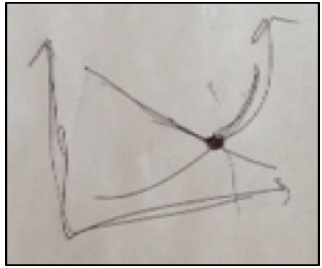

**Figure 2.** Students in one small group drew a diagram of their idea that thresholds occur when two curves cross one another.

**Implications for Seeing Class Concepts in Students' Everyday Lives**
Both authors have engaged a number of undergraduate STEM students in the activity of crafting a definition for *threshold*. The activity has potential for increasing student interest in physics. After engaging in the *threshold* activity, many students start seeing thresholds in their everyday lives. While there is a range, participants in our defining activities have said things like, "I see thresholds everywhere now!" or mentioned that their definition of threshold "bugs" them days later. Indeed, second author Atkins Elliott surveyed college STEM students in her pre-service teacher course one week after the in-class *threshold* defining activity. She asked, "Did you notice or think about thresholds in the past week?" 27 of 32 students, a relatively high percentage, responded affirmatively. This stands in contrast to traditional science courses where experiences of noticing or using course content outside of class are rare.[12] These survey results give preliminary evidence that the *threshold* activity can be used in supporting students to connect physics to their lives outside of the physics classroom.

**Demystifying Scientific Practice and Building Student Agency**
We posit that engaging students in crafting definitions from the ground up has the potential to impact student agency around definitions. One undergraduate physics major who experienced the *threshold* activity was asked to reflect on it. She noted that her experience in physics classes overall was one where she wasn't typically encouraged to explore her questions about definitions[9]:

> *Physics is, like, presented to you, like, this is how it is and you're never really encouraged to think about why certain distinctions are made and stuff…It [the threshold activity] was really interesting!...You don't really get to think about what it means for something to be something or not something...*

For students to approach definitions in science courses critically, i.e. to evaluate whether or not they understand them, whether or not they make sense in terms of their own conceptual frameworks, it helps for them to recognize that definitions are human constructions. Definitions are proposed, debated, and changed. Math education researchers Zandieh and Rasmussen[13] note the importance of involving students in "the mathematically deep issue of negotiating criteria for why certain elements…should or should not be part of the definition." They go onto say that the "significance of this type of activity for students is that it can contribute to developing a meta-understanding of mathematical definitions." If students know that there is this kind of social machinery going on "under the hood" of scientific definitions, they may be better positioned to interpret and use them appropriately, and to see themselves as the kind of person who can construct and improve definitions, too.

**Lack of Curriculum That Engages Students in Crafting Definitions**
Even in many reformed introductory physics curricula such as *Tutorials in Introductory Physics*[14] or Modeling Instruction[15,16] it is rare that students spend extended time crafting definitions for themselves. *Physics by Inquiry*[17] curriculum engages teachers in operationally defining concepts such as *density*. However, this curriculum is used almost entirely in teacher professional development and infrequently with introductory high school or college physics students.

One of our goals as physics educators is not only for students to understand scientific claims but to perform scientific practices as a way of developing and understanding those claims. This is a core element of the Next Generation Science Standards[18], which emphasizes "three-dimensional learning": engaging students in scientific practices to develop an understanding of disciplinary core ideas around concepts that cut across disciplines. Increasingly, we have noticed constructing and refining definitions as central to this practice — both as a scientific practice and key to understanding and using disciplinary core ideas.

**Conclusion**
The *threshold* defining activity, only 1-2 hours in length, has the strength of fitting into many different curriculum types across both high school and college contexts. It could be used as an end-of-year physics course activity to support students in seeing connections across content. It could be used at the beginning of a physics course as a way to begin demystifying definitions. Physics instructors could also use the *threshold* activity as scaffolding toward additional definitional activities for other key terms.

Our hope for sharing this activity is that it will aid physics educators in supporting students to see themselves as capable in a skill central to the practice of science.

**Acknowledgements**
The authors would like to thank Prof. Andrea diSessa and members of the Patterns Research Group at UC Berkeley. We gratefully acknowledge financial support by the Spencer Foundation (grant #201100101, A. diSessa, PI), and the National Science Foundation (grants #1638523 and #1638524, L. Atkins Elliot, PI). The opinions and results reported in this work are those of the authors, and are not necessarily endorsed by the foundations. We would also like to thank Dimitri Dounas-Frazer and Mel Sabella for their helpful feedback on the manuscript.

---

[1] https://www.tax.ny.gov/pubs_and_bulls/tg_bulletins/st/sandwiches.htm (2011), accessed March 18, 2017.

[2] If you would like to hear more about the New York sandwich debates, see this Sporkful Podcast episode where they debate whether or not hot dog counts as a sandwich: http://www.sporkful.com/john-hodgman-v-dan-pashman-are-hot-dogs-sandwiches/ (2015), accessed March 18, 2017.